# Non-invasive tracking of polarization rotation from speckle contrast measurement

Abhijit Roy,[1)] Rakesh Kumar Singh,[2)] and Maruthi M. Brundavanam[1)]

[1]*Department of Physics, Indian Institute of Technology Kharagpur, Kharagpur, West Bengal 721302, India*

[2]*Department of Physics, Indian Institute of Technology (Banaras Hindu University), Varanasi, Uttar Pradesh 221005, India*

*Email: bmmanoj@phy.iitkgp.ac.in*

A technique based on speckle contrast measurement is proposed and demonstrated to track the polarization rotation of an optical beam behind a scattering layer. It is shown that with the help of a reference speckle pattern with a known linear polarization, it is possible to track the polarization rotation behind a scattering layer in a non-invasive manner by measuring the contrast of the resultant speckle pattern. Working principle and experimental demonstration are presented, and results are compared with the theoretically known results in certain cases. Performance of the proposed technique is quantitatively evaluated by determining the polarization vector rotation behind the scattering layer for two different cases, namely polarization rotation by a half wave plate and sugar solution of different concentrations, and good match are found between the experimental results and theoretical prediction. Issues like resolution of determination of the polarization vector rotation and sugar solution concentration are also discussed in the context of the recovered results. The proposed technique can have potential application in non-invasive sensing and biomedical imaging.

Interference of coherent lights scattered by a random scattering medium gives rise to a complex pattern known as speckle [1]. The speckles can be characterized using different techniques such as intensity correlation, speckle contrast (SC) etc. [1]. Because of its simplicity, the intensity correlation technique is used widely in different applications such as looking through barrier [2], measuring viscosity of fluid [3], refractive index of a medium [4] etc. Investigating the spatial coherence and spatial polarization distribution of a speckle pattern, following intensity correlation, is found to be helpful in imaging [5] and determining vorticity of vortex beams from the speckle field [6].

The SC, another parameter which contains information about the spatial coherence-polarization property of the speckles, is reported to be a very effective tool in various applications. The measurement of SC is useful in characterizing different types of diffusive media [7]. Recently, a technique is proposed to determine the diffusion properties of a strong scattering medium from the SC, where the speckles are generated using a femtosecond laser pulse [8]. In another separate work, a relation between the SC and surface roughness is exploited to measure a wide range of surface roughness using a white light source [9]. Variation of SC with the point spread function of an imaging system provide the roughness and the correlation length of an object [10]. A linear relationship between the far-field SC and roughness of an illuminated surface can be exploited to measure roughness of different metallic surfaces [11]. In a separate study, a relation between SC and temporal degree of coherence of an illuminating source is established, and it is shown that the coherence length of the source can be determined from the SC [12]. The SC measurement can also be used in estimating low viscosity variation of fluids [13].

Another technique, known as the laser speckle contrast imaging (LSCI), is found to be very useful in biomedical applications [14]. It has been demonstrated that using a laser pointer-webcam setup and exploiting the LSCI technique, it is possible to image cerebral dynamic blood flow, inexpensively [15]. The viscosity and elasticity of tissues can also be determined by the LSCI method [16]. Separately, a technique called, Speckle Contrast Optical Tomography (SCOT), is introduced to image the distribution of three dimensional blood flow in deep tissues [17]. Recently, another non-invasive technique called, Speckle Contrast Optical Spectroscopy (SCOS), is proposed to probe microvascular blood flow in deep tissues [18]. It is also reported that different kinds of skin cancers can be detected and monitored from the SC measurement [19].

Contrast of the speckles, produced using a white light source, varies with the distance near the image plane, and the variation strongly depends on the mean wavelength, temporal coherence length of the source and the surface roughness of the medium [20]. Laser speckle contrast reduction by the multimode optical fiber bundle with combined temporal, spatial and angular diversity has been applied for imaging application [21]. The contrast of time-averaged speckle pattern, formed at a defocused image plane due to a vibrating diffused object, is observed to be reducing with the increase in the rate of vibration [22]. Techniques based on superposition of speckle patterns and modulation of polarization of the input beam have also been proposed to control the SC [23, 24]. Investigations in Refs. [12] and [24] establish a relationship of SC with the input polarization and the spatial polarization distribution of the speckle pattern, which can be exploited for polarization related study through a random scattering medium.

Different techniques have been proposed to examine the propagation of polarized field through random scattering media. A technique, based on cross-correlation of four speckle patterns, is proposed to retrieve the Stokes vector of the input beam through a thick scattering medium [25]. Retrieval of complex coherence function is also reported to be used to recover the polarizations of the input objects through a random birefringent scattering medium [26]. Polarization recovery through a bulk scattering medium is possible by modulating the shape of the input wavefront [27]. Recently, different techniques based on four shot measurement and off-axis holographic approach have been proposed to retrieve different polarization vectors through a scattering layer [28, 29]. Although the techniques mentioned in Refs. [25-29] are proposed for polarization retrieval, no significant attempts seem to exist to track the polarization rotation through a random scattering medium in real time or from a single shot intensity measurement except in Ref. [30] wherein a technique based on intensity cross-correlation is proposed to accurately track the rotation of polarization vector through a scattering medium.

In this letter, we propose an alternative and a new technique, based on superposition of two speckle patterns to track the rotation of polarization vector through a scattering layer in real time, using the prior knowledge of polarization of one of the two superposing speckle patterns. In contrast to the technique presented by Freund et al. [30], we make use of an independent reference speckle pattern with a fixed polarization orientation, and this speckle pattern is generated from a completely independent scatterer. Therefore, this provides an opportunity to develop a single shot approach to track the polarization orientation of a desired field in real time. The potentiality of the proposed technique in sensing application is demonstrated by successfully recovering the concentration of sugar solution through a scattering layer. The theoretical details along with the experimental demonstration are presented in the following sections.

Let us consider that the propagation of a monochromatic, linearly polarized sample field (SF) with the polarization vector oriented at an angle $\phi$ with respect to the x-axis through a scattering layer results in a sample random field (SRF), $\mathbf{E_S}(\mathbf{r}, t)$, at the observation plane. As a scattering layer does not change the polarization of the input field, $\mathbf{E_S}(\mathbf{r}, t)$ has the same polarization as the input SF. The SRF, $\mathbf{E_S}(\mathbf{r}, t)$, at the transverse observation plane, $\mathbf{r}$ can be written as

$$\mathbf{E_S}(\mathbf{r}, t) = E_S(\mathbf{r}, t) \cos\phi\, \hat{\mathbf{x}} + E_S(\mathbf{r}, t) \sin\phi\, \hat{\mathbf{y}} \tag{1}$$

where $\hat{\mathbf{x}}$ and $\hat{\mathbf{y}}$ are the orthogonal unit vectors, and $E_S(\mathbf{r}, t)$ is the magnitude of the SRF.

The contrast, C of intensity distribution of the SRF can be determined from the standard deviation of intensity, $\sigma_I$ and the mean intensity, $\langle I \rangle$ following [1]

$$C = \frac{\sigma_I}{\langle I \rangle} = \frac{\sqrt{\langle I^2 \rangle - \langle I \rangle^2}}{\langle I \rangle} \tag{2}$$

The SC of any spatially uniformly polarized speckle pattern is equal to unity [1] which can also be calculated from Eq. (2) using Eq. (1). In the present case, with change of $\phi$, as the speckle pattern still remains spatially uniformly polarized, the SC becomes independent of $\phi$, and is always equal to unity. Hence, from the study of SC, it is not possible to determine the orientation of polarization vector of the SF. In order to determine $\phi$, in the present work, the SRF, $\mathbf{E_S}(\mathbf{r}, t)$ is superposed with another spatially uniformly polarized random field, $\mathbf{E_R}(\mathbf{r}, t)$, referred as reference random field with known polarization. Assuming that the reference random field vector makes an angle $\theta$ with the x-axis, the $\mathbf{E_R}(\mathbf{r}, t)$ is written as

$$\mathbf{E_R}(\mathbf{r}, t) = E_R(\mathbf{r}, t) \cos\theta\, \hat{\mathbf{x}} + E_R(\mathbf{r}, t) \sin\theta\, \hat{\mathbf{y}} \tag{3}$$

where $E_R(\mathbf{r}, t)$ is the magnitude of the reference random field. The average intensity of the superposed random field can be calculated as

$$\langle I(\mathbf{r}) \rangle = \langle I_S(\mathbf{r}) \rangle + \langle I_R(\mathbf{r}) \rangle + [\Gamma_{RS}(\mathbf{r},\mathbf{r}) + \Gamma_{SR}(\mathbf{r},\mathbf{r})]\cos(\theta - \phi) \tag{4}$$

where $\Gamma_{RS}(\mathbf{r},\mathbf{r})$ and $\Gamma_{SR}(\mathbf{r},\mathbf{r})$ are the mutual coherence functions of the sample and reference random fields, and are defined as $\Gamma_{ij}(\mathbf{r},\mathbf{r}) = \langle E_i^*(\mathbf{r}) E_j(\mathbf{r}) \rangle$. As the two random fields are experimentally generated from two independent scattering media, following [26] $\Gamma_{RS}(\mathbf{r},\mathbf{r})$ and $\Gamma_{SR}(\mathbf{r},\mathbf{r})$ are assumed to be zero, and under this assumption, the following calculations are done.

$$\langle I(\mathbf{r}) \rangle = \langle I_S(\mathbf{r}) \rangle + \langle I_R(\mathbf{r}) \rangle \tag{5}$$
$$\langle I^2(\mathbf{r}) \rangle = \langle I_S^2(\mathbf{r}) \rangle + \langle I_R^2(\mathbf{r}) \rangle + 2\langle I_S(\mathbf{r}) \rangle \langle I_R(\mathbf{r}) \rangle + 2\langle I_S(\mathbf{r}) \rangle \langle I_R(\mathbf{r}) \rangle \cos^2(\theta - \phi) \tag{6}$$

The contrast of the superposed intensity distribution is calculated using Eqs. (5) and (6), and can be found as

$$C = \frac{\sqrt{\langle I^2 \rangle - \langle I \rangle^2}}{\langle I \rangle} = \frac{\sqrt{\langle I_S(r) \rangle^2 + \langle I_R(r) \rangle^2 + 2 \langle I_S(r) \rangle \langle I_R(r) \rangle \cos^2(\theta - \phi)}}{\langle I_S(r) \rangle + \langle I_R(r) \rangle} \quad (7)$$

If the spatial average intensity of the sample and reference random fields are made same i.e. $\langle I_S(r) \rangle = \langle I_R(r) \rangle$, Eq. (7) is modified as

$$C = \sqrt{\frac{1 + \cos^2(\theta - \phi)}{2}} \quad (8)$$

It can be observed from Eq. (8) that the contrast varies from 0.707 to 1.0 in a sinusoidal fashion with the change of the angle difference between the polarization vectors of the two linearly polarized random fields. The unknown orientation of the polarization vector of SF ($\phi$) can be estimated from the determined contrast, C of the superposed intensity distribution using the prior knowledge of orientation of the polarization vector of reference random field ($\theta$).

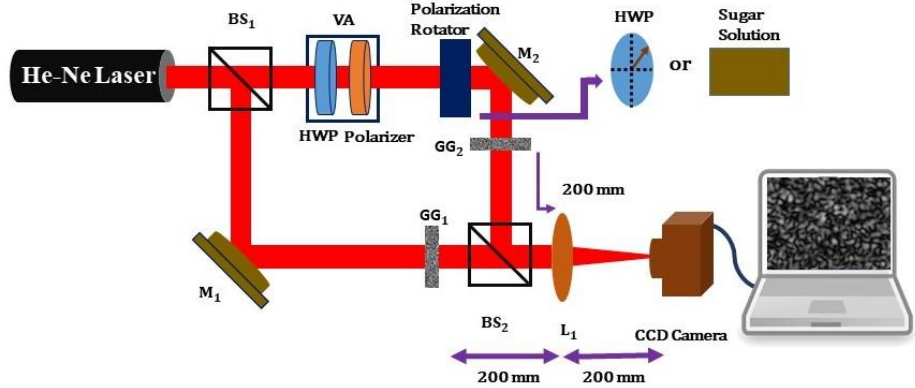

FIG. 1. Schematic diagram of the experimental setup.

The schematic of the experimental setup is presented in Fig. 1. A horizontally polarized beam of light from a He-Ne laser source of wavelength 632.8 nm enters a Mach-Zehnder interferometer formed by mirrors $M_1$, $M_2$ and non-polarizing beam splitters $BS_1$, $BS_2$. The light reflected from beam splitter, $BS_1$ is folded by mirror $M_1$ and is passed through a ground glass (GG) plate, $GG_1$. The speckles generated from $GG_1$ form the reference arm of the interferometer and are referred as reference speckles. On the other hand, the light transmitted through the beam splitter $BS_1$ is passed through a polarization rotator which rotates the polarization vector of the light. The light transmitted through the polarization rotator is then folded by mirror $M_2$ and is passed through another GG plate, $GG_2$. The speckles generated from $GG_2$ constitute the sample arm of the interferometer and are referred as sample speckles. As the GG plates do not scramble the polarization of input beam, the generated speckles will have the same polarization as the input beam. Here, in our case, the reference speckles are horizontally polarized and hence, in Eq. (8) $\theta = 0^0$. Whereas the polarization of sample speckles changes depending on orientation of fast-axis of a half-wave plate (HWP) or due to test solution like sugar, which are used as polarization rotators in the present study, as shown in Fig. 1.

The sample and reference speckles are superposed using BS$_2$, and the superposed speckles are recorded using a CCD camera which is kept at the rear focal plane of a Fourier transforming lens, L$_1$ of focal length of 200 mm. And both the GG plates are kept at the front focal plane of L$_1$. A variable attenuator, VA which consist of another half-wave plate and a polarizer with pass axis along the horizontal direction is placed before the polarization rotator to control the average intensity of the sample speckles in order to satisfy the condition mentioned before Eq. (8). At first, the HWP is used as the polarization rotator, and the fast-axis of the HWP is rotated from $0^0$ to $180^0$ in steps of $3^0$, and the superposed speckles are recorded by the CCD camera. Both the sample and reference speckles are fully developed at the camera plane, and follow Gaussian statistics.

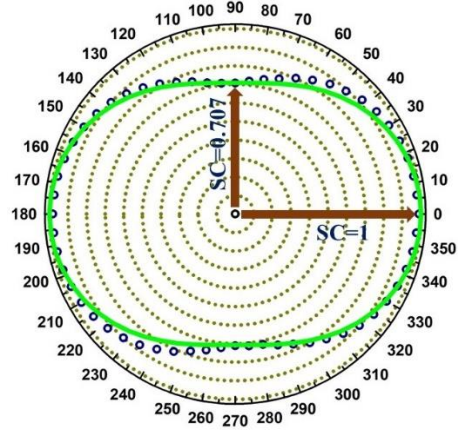

FIG. 2. Change of SC with the orientation of polarization vector of the SF ($\phi$).

The contrast of a recorded speckle pattern is determined using Eq. (2), where the spatial average is taken over different spatial points on the speckle pattern. The SCs of both the reference and sample speckle pattern are found to be 1. In case of superposed speckle pattern, the estimated SC is found to be changing with the orientation of polarization vector of the SF ($\phi$), and the variation is presented in Fig. 2. The circles are the experimental results, whereas the solid line is the theoretically predicted variation following Eq. (8) with $\theta = 0^0$, and it can be observed that the experimental results are matching well with the theoretical prediction. The variation of SC with $\phi$ confirms that the proposed technique is able to sense the polarization rotation behind a scattering layer.

In order to confirm that exploiting the proposed technique, exact orientation of the polarization vector behind a scattering layer can also be retrieved, which is required to accurately track the rotation of polarization vector, the following experiment is performed. The fast-axis of the HWP is kept at an angle of $15^0$ and $30^0$ with respect to the x-axis, which results in change in orientation of the polarization vector of sample speckles to $30^0$ and $60^0$, respectively, and the superposed speckles are recorded. The SCs of the superposed speckle patterns are determined, and the orientations of polarization vectors of the SF are calculated from the estimated SC following Eq. (8) using the knowledge of the polarization of reference speckles ($\theta = 0^0$), and the results are presented in Table 1.

Table 1. Retrieved orientations of the polarization vector of sample field

| Original Orientation | Expected SC | Retrieved SC (Standard Deviation) | Retrieved Orientation (Standard Deviation) |
|---|---|---|---|
| $30^0$ | 0.9354 | 0.9351 (0.0009) | $30.08^0$ ($0.22^0$) |
| $60^0$ | 0.7906 | 0.7898 (0.0003) | $60.16^0$ ($0.05^0$) |

The retrieved SC and the orientations of polarization vectors of the SF, presented in Table 1, are averaged over 21 sets of speckle patterns, and it is observed that the experimental results are matching well with the actual values. The standard deviations in the experimental results, presented in the parenthesis in Table 1, are observed possibly due to the noise level of CCD camera and power fluctuation of the laser. In order to determine the resolution of tracking the polarization rotation, employing the proposed technique, the fast-axis of the HWP is rotated from $15^0$ to $16.5^0$ in steps of $0.5^0$, and the orientations of the polarization vectors are retrieved from the SC of the superposed speckle patterns following the above mentioned method. The results are presented in Table 2, where it can be found that the observed maximum standard deviation is $0.72^0$ which restricts the maximum achievable resolution approximately to $1^0$. However, the resolution of tracking can be improved using CCD camera of lower noise level and laser with higher stability in power fluctuation.

Table 2. Tracking of polarization rotation

| Original Orientation | Retrieved SC (Standard Deviation) | Retrieved Orientation (Standard Deviation) |
|---|---|---|
| $30^0$ | 0.9351 (0.0009) | $30.08^0$ ($0.22^0$) |
| $31^0$ | 0.9312 (0.0005) | $31.02^0$ ($0.11^0$) |
| $32^0$ | 0.9266 (0.003) | $32.12^0$ ($0.72^0$) |
| $33^0$ | 0.9213 (0.0007) | $33.35^0$ ($0.17^0$) |

Different sensing technologies have also been developed employing the speckles. Controlling the scattering of light from Au nanorods embedded in sol-gel host by electrical means opens a new way of developing different electro-optical devices for optical modulation [31]. Remote fingerprint sensor based on temporal tracking of speckle pattern is developed recently [32]. In the present study, the potentiality of the proposed technique in sensing application is demonstrated by measuring concentration of sugar solution through a random scattering medium. The polarization vector of a linearly polarized light gets rotated while passing through a sugar solution due to the circular birefringence of sucrose, and the amount of rotation is written as $\phi = S \times L \times D$, where D is the concentration of sugar in the solution, L is the length of the container and S is the optical activity of the sugar solution. In the present case, the investigation is done by replacing the HWP in the sample arm of the interferometer with a container of length of 4.4 cm, and the concentration, D of sugar solution in the container is changed as follows: 0 g/ml, 0.6 g/ml, 0.8 g/ml, and 1.0 g/ml. The optical activity, S of the sugar solution is determined separately employing the traditional polarimetry experiment, and is found to be 41.67 deg mL $g^{-1}$ $dm^{-1}$. The superposed speckles for different concentrations of sugar solution are recorded, and the SC for each concentration is averaged over 21 sets of speckle patterns. The orientation of the polarization vector, $\phi$ is estimated using Eq. (8) from the average SC, and the concentration, D is calculated following $D = \frac{\phi}{S \times L}$, using the earlier determined value of S. The retrieved and original concentrations are presented in Table 3, where it can be observed that the retrieved concentrations are matching well with the original values. As the noise level of camera and laser power fluctuation introduces instability in the fourth decimal of SC, the observed instability in the third decimal of SC, in case of sugar solution, is possibly due to the motion of sugar particles in the solution which introduces temporal fluctuation to the speckles and recorded intensity. Hence, change in concentration, which introduces change in the second decimal of the SC, can be resolved following the proposed technique, and in the present case, it is found to be 0.2 g/ml. Moreover, it can be observed that, initially, sugar solution of concentration 0.6 g/ml is required to change the SC from 1.0 (when the concentration in 0 g/ml) to 0.99, which restricts the minimum retrievable concentration to 0.6 g/ml. As initially, 0.6 g/ml is required to introduce change in the second decimal of the SC, the observed standard deviation in this case is high compared to the mentioned resolution. Besides, as S increases with decrease in the wavelength of light, less concentration can introduce the required change in SC if source of lower wavelength is used, this in turn can improve the minimum retrievable concentration and resolution of the concentration measurement.

Table 3. Retrieved concentrations of sugar solution

| Original Concentration in g/ml | Retrieved SC (Standard Deviation) | Retrieved Angle (Standard Deviation) | Retrieved Concentration in g/ml (Standard Deviation) |
|---|---|---|---|
| 0.6 | 0.9901 (0.0109) | $10.91^0$ ($6.03^0$) | 0.59 (0.33) |
| 0.8 | 0.9818 (0.0077) | $15.24^0$ ($3.41^0$) | 0.83 (0.19) |
| 1.0 | 0.9756 (0.0072) | $17.89^0$ ($2.84^0$) | 0.98 (0.15) |

In conclusion, we have proposed and demonstrated a technique based on speckle interference to sense the polarization rotation behind a scattering layer using a single speckle pattern. We have demonstrated the ability of the proposed technique to perfectly track the polarization rotation in real time with a resolution of $1^0$ in a non-invasive manner. The resolution of tracking can be improved using detector having lower noise level, higher dynamic range, and laser with higher stability in power fluctuation. The successful recovery of the concentration of sugar solution, hidden behind the scattering medium, with resolution of 0.2 g/ml highlights potentiality of the proposed technique in many other practical applications. Moreover, this approach can be extended to simultaneously detect polarization rotation of the electric fields at different spatial points, i.e. wide field rotation measurement. The limitation of the proposed technique is that perfect tracking and retrieval of polarization is possible for $0^0 \leq \phi \leq 90^0$ due to $\cos^2(\theta - \phi)$ in Eq. (8), and for other angles, it will be the repetition of the values for $0^0 \leq \phi \leq 90^0$. Whereas, apart from non-invasiveness, another major advantage of the technique lies in the fact that a completely independent scattering medium can be used in the reference arm for different applications.

We acknowledge support from SERB-DST, Government of India (Project Nos. SR/FTP/PS-170/2013 and EMR/2015/001613), Indian Institute of Technology Kharagpur (Project No.

IIT/SRIC/PHY/VBC/2014-15/43) and Indian Institute of Technology (Banaras Hindu University) (Project No. Plan-OH-35 R & D grant).